\documentclass{aa}
\usepackage{graphics}
\begin{document}
\thesaurus{11        
          (11.09.1;  
	   11.19.3;  
           11.16.1)}  
\title{Star formation along a misaligned bar in the peculiar starburst galaxy Mkn 439}
\author{Aparna Chitre, U.C. Joshi and S.Ganesh}
\offprints{A. Chitre}
\institute{Physical Research Laboratory, Navrangpura, Ahmedabad 380 009, INDIA.
\\email: chitre@prl.ernet.in}
\date{Received ; accepted}
\authorrunning {A.Chitre}
\titlerunning {Star formation in Mkn 439}
\maketitle

\begin{abstract}
We report the detection of massive star formation along a bar in the peculiar
starburst galaxy Mkn 439. We present optical $B$, $R$ and $H\alpha$+$[NII]$ emission
line images as well
as $H$ band images to show that the signature of the bar becomes progressively
weak at longer wavelengths. Moreover, this bar is misaligned with the main
body of the galaxy.
The peak $H\alpha$ emission does not coincide with the bluest regions seen
in the colour maps. We infer that the starburst is young since the stars in the
burst have not started influencing the light in the near infrared. There are
indications of dust in the inner regions of this galaxy.
\keywords{Galaxies : individual : Mkn 439 -- Galaxies : starburst -- Galaxies : photometry}
\end{abstract}
\section{Introduction}
Mkn 439 (NGC 4369, UGC 7489, IRAS 12221+3939) is a nearby early type
starburst galaxy (z=0.0035).
It has been
classified as a starburst by Balzano (\cite{balz}) based on the
equivalent width of $H\alpha$ and also belongs to the
$\it IRAS$ Bright Galaxy Sample (Soifer et al. \cite{soif}).
On the basis of multiaperture near
infrared photometry and optical spectroscopy, Devereux (\cite{dev}) describes
this galaxy as a M82 type starburst galaxy. Rudnick \& Rix (\cite{rudrix})
report an azimuthal asymmetry in the stellar mass distribution of Mkn 439
based on the $R$ band surface brightness.
The peculiar morphology of Mkn 439 attracted our attention during
the course of an optical imaging study of a sample of starburst
galaxies derived from the Markarian lists (Chitre \cite{chitre}).
The galaxy image was nearly circular
and appeared featureless in long exposure images. The outer isophotes
were smooth and
nearly circular in $B$ and $R$ bands. However, the isophotal
contours show highly complex features in the inner parts. Moreover, the
strength of these features is wavelength dependent.
Wiklind \& Henkel (\cite{wik})
report the detection of a molecular bar in the central region of this galaxy
based on the results of CO mapping.
No detailed surface photometric studies of this galaxy have been reported.
Usui, Saito \& Tomita (\cite{usui}) report the detection of two regions bright
in $H\alpha$ that are displaced from the nucleus and faint
emission from the nucleus.
However, their data were obtained at a seeing of 5\arcsec. In order to
study the spatial distribution of various stellar populations in Mkn 439,
we imaged this galaxy in $B$, $R$, $H\alpha$ and $H$ bands. The $B$ and
$R$ band continuum trace the intermediate age populations while $H\alpha$
traces the young, massive stellar populations. The
infrared continuum
of galaxies is dominated by evolved stellar populations. 
Hence,
the $H$ band and the line emission images can be used alongwith
the optical continuum images
to separate the young and old stellar populations spatially.
\section{Observations and data reduction}
\subsection{Optical ($B$,$R$) and $H\alpha$ imaging}
The $B$, $R$ and $H\alpha$ images were obtained under photometric conditions
from the 1.2m telescope at Gurushikhar, Mt. Abu. The
images were taken at the Cassegrain focus employing a thinned
back illuminated Tektronix 1K $\times$ 1K CCD. Binning of 2 $\times$ 2 was
employed before recording the images to
increase the signal-to-noise ratio of the measurements and keeping in mind the
data storage requirements.
The final resolution was 0.634$^\prime$$^\prime$/pixel which is sufficient to
sample the point spread function (PSF)
appropriately. Typical seeing (full width at half maximum (FWHM) of
the stellar images) was $\sim$
1.8$^\prime$$^\prime$ for the images.
 For the $H\alpha$ images a filter
having FWHM of 80 \AA was used. Another off-band filter of the same
FWHM was used to measure the galactic red
continuum.
About 3-4 exposures were taken in each of the photometric bands.
The total exposure times were 510 sec, 360 sec and 1600 sec in $B$,
$R$ and $H\alpha$ respectively.
Standard stars from Landolt (\cite{land}) were observed to calibrate the broad band data. Twilight flats were taken and median filtered to construct the
master flats. The data was reduced using IRAF \footnote{IRAF is
distributed by National Optical Astronomy
Observatories, which is operated by the Association of Universities Inc.
(AURA) under cooperative agreement with the National Science Foundation,
USA.}
on the IBM-6000 RISC
at PRL, Ahmedabad. A detailed reduction procedure can be found in
Chitre \& Joshi (\cite{cucj}).
\subsection{$H$ band images}
The $H$ band images were recorded with a 256$\times$256 NICMOS array
at the 1.2 m Gurushikhar telescope. An exposure of 30 sec ensured that
the background and the galaxy signal were in the linear portion of the
detector response
curve
(Joshi et al. \cite{jetal}). 
Observations were made by alternating between the galaxy and positions 4\arcmin-5\arcmin to the north and south till a total integration time of 600 seconds
on the galaxy was achieved.
Several dark frames having the same time sequences as that of galaxy or sky were
 taken and  median filtered master dark frames were constructed.
The median filtered master sky frames were constructed using several sky frames
with integration times equal to those given for the galaxy. All the source
frames were corrected for the sky background by subtracting the master sky frame
from the source frames. As the program galaxy does not occupy the whole
detector array, the residual sky was determined from the image corners and
the images were then corrected for residual sky.
The dark subtracted sky frame was used to construct the master flat.
The sky corrected galaxy frames were corrected for flat field response of the
detector by dividing the galaxy frames by the master flat.\\
        Finally, the galaxy images
were aligned by finding the center of the galaxy nucleus using the IMCNTR task
in IRAF and co-added to improve the S/N ratio. The plate scale was selected to
be 0$^\prime$$^\prime$.5 per pixel. Faint standard stars from the UKIRT
lists were observed for calibration.
\begin{table}\caption[]{Global properties of Mkn 439}
\begin{tabular}{ll}
\hline
\hline
Parameter&Value\\
\hline
$\alpha$(2000)&12$^h$22$^m$0.8$^s$.4\\
$\delta$(2000)&39$\degr$39$^{\arcmin}$41$^{\arcsec}$\\
RC3 type&RSAT1\\
UGC type& S0/Sa\\
$^{\mathrm{b}}$Adopted distance &18 Mpc\\
$^{\mathrm{a}}$$B^{^0}_T$ &12.27\\
$^{\mathrm{a}}$$(U-B)_T$&-0.02\\
$^{\mathrm{a}}$$(B-V)_T$&0.65\\
L$_{FIR}$&4$\times$$10^9$ $L_\odot$\\
\hline
\end{tabular}
\begin{list}{}{}
\item[$^{\mathrm{a}}$] RC3
\item[$^{\mathrm{b}}$] Deutsch \& Willner (\cite{deut})
\end{list}
\end{table}
\begin{figure*}
\resizebox{\hsize}{!}{\includegraphics{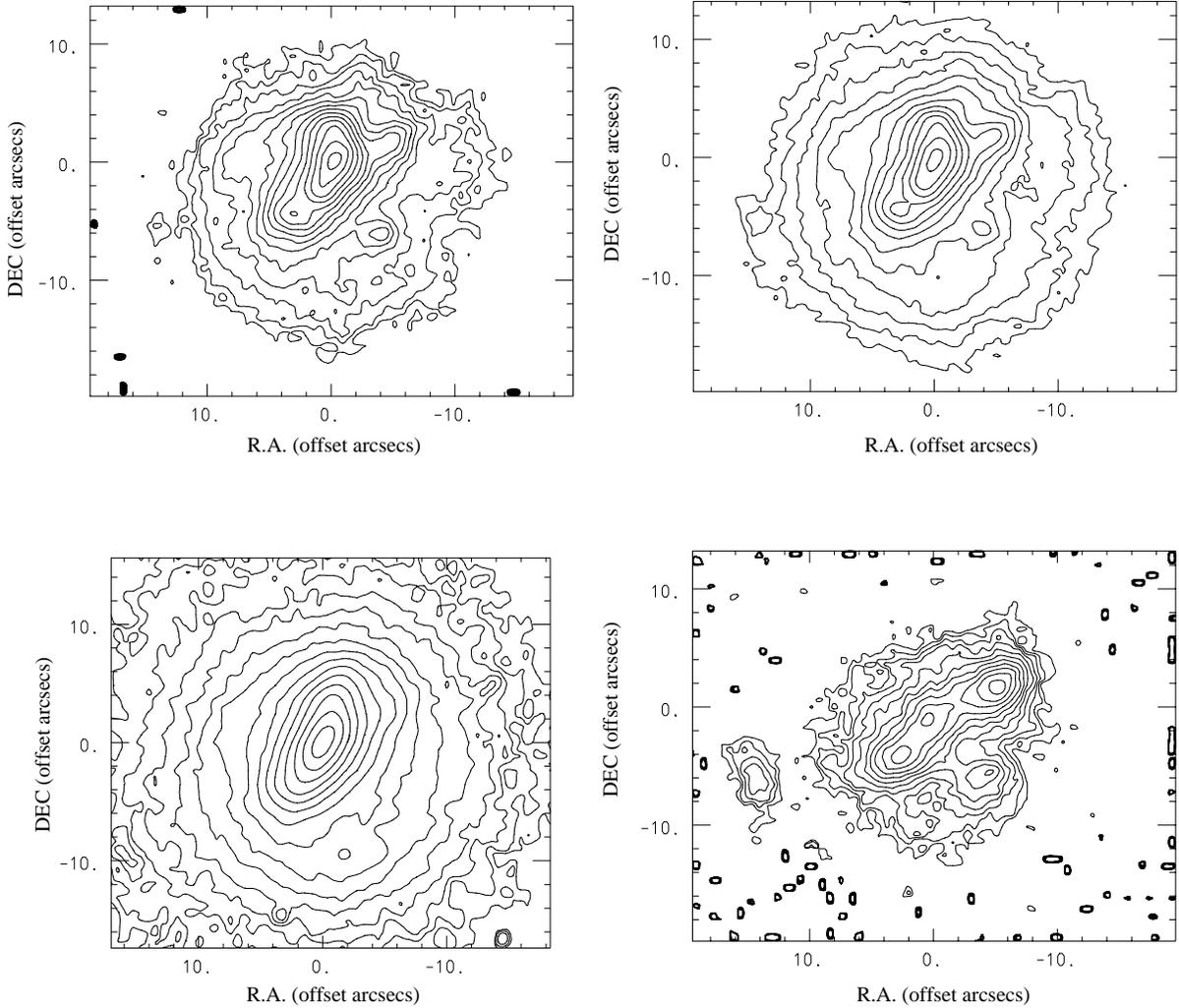}}
\caption{Top panel: $B$ (left) $R$ (right) band contours. Bottom panel:  $H$ (left) and continuum subtracted $H\alpha$+$[NII]$ (right) isophotal contours of Mkn 439. North is at the top and East is to the left. The isophotal contours are plotted on a logarithmic scale at intervals of 0$^m$.3 with the peak contours being at 17$^m$.5, 17$^m$.0 and 14$^m$.7 for $B$, $R$ and $H$ respectively. The $H\alpha$+$[NII]$ contours are plotted such that the lowest contour corresponds to 2$\sigma$ of the background.}
\label{cont}
\end{figure*}
\section{Morphology of Mkn 439}
\begin{figure}
\resizebox{\hsize}{!}{\includegraphics{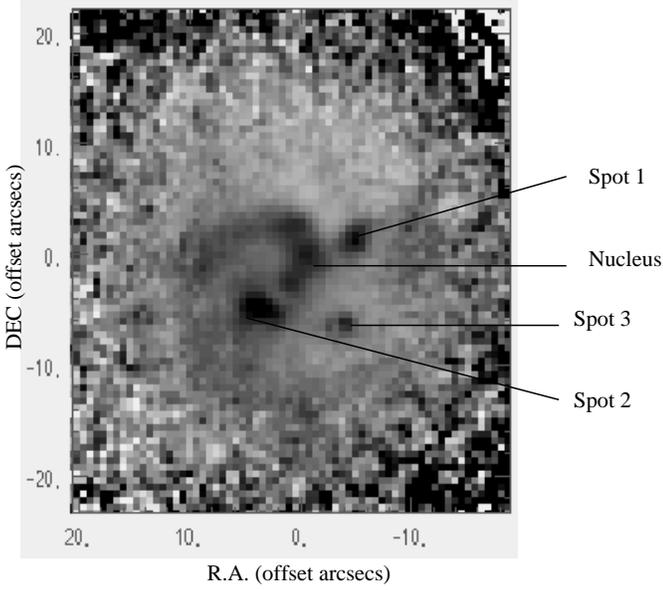}}
\caption{$B$-$H$ colour map of Mkn 439}
\label{bhcol}
\end{figure}

Fig. \ref{cont} illustrates the isophotal contours of the inner
25\arcsec\ of Mkn 439 in $B$, $R$,
continuum subtracted $H\alpha$ and $H$ band. A comparison of the various
panels in Fig. \ref{cont} shows that the morphological structures vary
at different wavelengths. The morphology of Mkn 439
in the $B$ band is
characterized by smooth outer isophotes and a
very complex light distribution in the inner region.
The
central region is elliptical and is elongated in the NS direction.
Faint indications of a spiral arm in the NE direction
is seen in the isophotal maps in $B$ and $R$. The contour maps show
two projections - one along the NW and the other along the SE from the
nuclear region. These projections are most prominent in the $B$ continuum,
getting progressively fainter at longer wavelengths and
nearly disappearing in $H$. 
The $B$ band image shows another condensation to the
SW of the nuclear region. Similar to the projections, this feature also
gets progressively fainter at longer wavelength.
The $H$ band image
shows smoother isophotes. The signature of the projections is absent at this
wavelength. As seen in the $R$ band, the outer isophotes are nearly circular.
However, unlike other optical bands, there are no spurs or bar-like features
apparent in the $H$ band image.
The continuum subtracted $H\alpha$ image shows an elongated bar-like strucure
corresponding
to the projections seen in the contour maps.
$H\alpha$ emission is seen along
the bar in the form of clumps. Emission is most intense at the ends of the bar,
though it is found to extend throughout the body of the galaxy. Emission
from the nucleus is
much fainter as compared to that from the clumps in the
bar ends. $H\alpha$ emission is maximum in Spot 1. 
The clump
of $H\alpha$ emission seen to the E of the extended emission has no
counterpart in the continuum colour map. The bright blobs of emission
in $H\alpha$ have no counterparts in the $H$ band. This indicates that
the HII regions are young and have not yet evolved enough to form a
considerable number of red giants and supergiants to start influencing
the light in the $H$ band. It is also seen that the latest episode of star
formation is misaligned with the isophotal contours of the near infrared
continuum.
The ($B$-$H$) colour map (Fig. \ref{bhcol}) was constructed by scaling
the images, rotating and aligning them. It shows interesting features. 
A bar-like structure made up of blue clumps is seen in the central part of
the galaxy. A spiral arm starts from the nuclear region and curves towards the
eastern side. A distinct blue clump is present at either end of the bar
marked as Spot 1 and Spot 2 in Fig. \ref{bhcol}. These correspond to the
ends of the two projections seen in the isophotal contours in $B$.
Another blue region (Spot 3) is seen about 8\arcsec\ to the south of Spot 1.
The clump of $H\alpha$ emission seen to the E of the extended emission has no
counterpart in the continuum colour map. The ($B$-$R$) and ($B$-$H$) colours of these
regions are listed in Table 2.
The isophotal contours of Mkn 439 appear different in the optical, near
infrared and the line emission indicating the spatial separation of the
distribution of these various populations. The gaseous component in the galaxy
appears to be under the influence of a potential which has distributed it
in the form of a gaseous bar. Compression of the gas in the bar has led to
the formation of young, massive stars which are seen as clumpy HII regions
along the bar. We infer that the latest dynamical episode experienced by the
galaxy has given rise to the formation of young, massive stars along the bar
as a result of the response of the gas to the perturbing potential. A comparison of the $H\alpha$ contours in Fig. \ref{cont}
and Fig. \ref{bhcol} reveals that no HII regions are seen in the blue spiral
arm-like feature emerging from the nucleus indicating that the blue spiral arm
is made up of an intermediate age stellar population.
Wiklind \& Henkel (\cite{wik})
report the detection of a molecular bar in the central region in this galaxy
based on CO mapping. They observed Mkn 439 in both the J=1-0 and J=2-1
line of CO and found that the ratio of the J=2-1 to the J=1-0 intensity
varies with position and inferred that this was due to changing physical
conditions in the molecular cloud population. The contour maps of these
two transitions can be found in Wiklind (\cite{wikthes}).
Many galaxies with weak stellar bars have been
found to contain pronounced bar-like gas distributions similar to the one
found in Mkn 439. For example, the center of the nearby Scd galaxy IC 342
harbors a bar-like molecular gas structure and a
modest nuclear starburst (Lo et al. \cite{lo}; Ishizuki
et al. \cite{ish}). Other examples of galaxies having a molecular bar at
their centers are NGC 253 and M83. Simulations by Combes (\cite{combes})
describe the formation of a gas bar which is phase shifted from the
stellar component in the innermost regions of a galaxy due to the existence
of perpendicular orbits. However, her models describe the situation for the
nuclear bars and in the innermost 1kpc region. An alternative explanation
could be that two unequal mass spirals have merged to form the S0 galaxy.
Bekki Kenji (\cite{bek}) suggests that S0 galaxies are formed by the merging of
spirals and when two spirals are of unequal mass, the S0 galaxy thus formed
has an outer diffuse stellar envelope or a diffuse disk like component
and a central thin stellar
bar composed mainly of new stars.
\begin{figure}
\resizebox{\hsize}{!}{\includegraphics{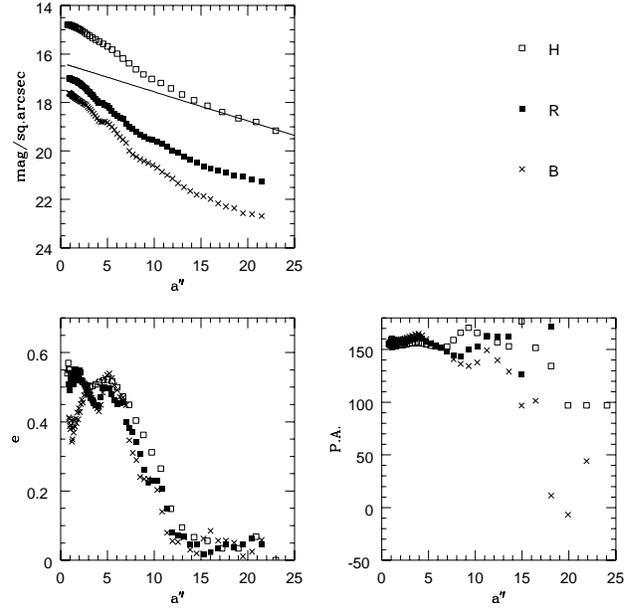}}
\caption{The surface brightness, ellipticity (e) and position angle (P.A.)
profiles of Mkn 439. The fitted disk in the $H$ band is shown by the line.}
\label{fit}
\end{figure}
\begin{figure}
\resizebox{\hsize}{!}{\includegraphics{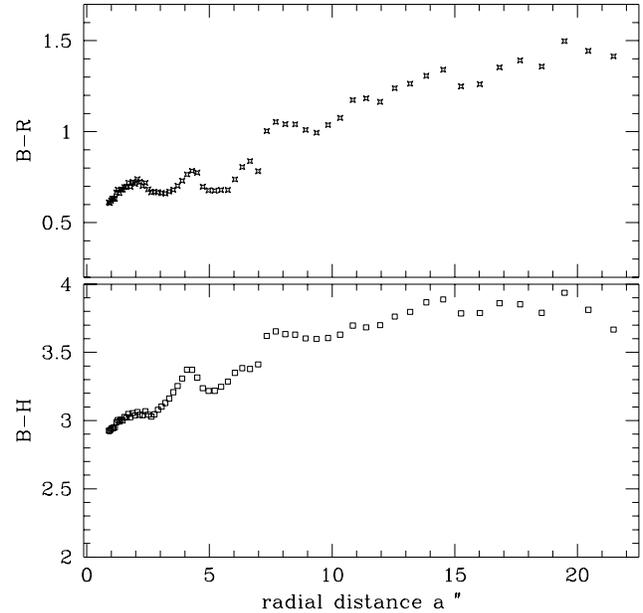}}
\caption{Radial distribution of $B$-$R$ and $B$-$H$ colours}
\label{brh}
\end{figure}

\begin{table}\caption[]{Colours of clumpy regions}
\begin{tabular}{lll}
\hline
\hline
region&$B$-$R$&$B$-$H$\\
\hline
nucleus&0.4&2.8\\
spot 1&0.7&2.9\\
spot 2&0.6&2.6\\
spot 3 &0.8&3.2\\
\hline
\end{tabular}
\end{table}
\section{Isophotal analysis}
In order to provide a quantitative description of the morphological
aspects at various wavelengths,
we explored Mkn 439 using ellipse fitting techniques.
The procedure consists of fitting
elliptical isophotes to the galaxy images and deriving 1-dimensional azimuthally
averaged radial profiles for the surface brightness, ellipticity and
the position angle
based on the algorithm given by Jedrejewski (\cite{jedr}). This technique
has been used successfully in studying various structures in galaxies like
bars, rings, shells, etc. and in searching for dust in them (Bender \& M\"{o}llenhoff
\cite{bend}; Wozniak et al. \cite{woz}
and Jungweirt, Combes \& Axon \cite{jung}). Multiband isophotal analysis
can also be
used to indicate whether the reddening seen in colour maps is due to a
redder stellar population or due to the presence of dust
(Prieto et al. \cite{prietoa}, \cite{prietob}).
The surface brightness distribution and the variation of the position angle
and ellipticity of the isophotes in each filter (Fig. \ref{fit})
were obtained by fitting ellipses to the images in each filter using the
ISOPHOTE package within STSDAS\footnote{The Space Telescope Science Data
Analysis System STSDAS is distributed by the Space Telescope Science
Institute.}. The detailed fitting procedure used is outlined in
Chitre (\cite{chitre}). The radial distribution of the colour
indices (Fig. \ref{brh}) was derived from the surface
brightness profiles.
Fitting isophotes to the images reveals changing ellipticity and position
angle throughout the body of the galaxy (refer Fig \ref{fit}).
The luminosity profile is smooth
except for small features at 5\arcsec\ and 10\arcsec\ in the optical bands.
An inspection of Fig. \ref{brh} shows that the galaxy is bluest near the
center and gets redder outwards. The ellipticity of the elliptical feature is maximum at the center and goes on decreasing outwards unlike a bar in which ellipticity increases accompanied by a constant position angle. The ellipticity profile shows a double
peaked structure in the inner region. The first peak is seen between
2\arcsec- 3\arcsec\ and the second peak at 5\arcsec. The
ellipticity of the first peak is wavelength dependent, the isophotes at
shorter wavelengths being rounder. The colour map also shows a small
local redder region between 3\arcsec\ and 4\arcsec.
The surface brightness profiles also show a small dip
in the intensity at shorter wavelengths at 4\arcsec. All these features
indicate the presence of dust in the inner 4\arcsec\ of this galaxy.
van den Bergh \& Pierce (\cite{van}) do not find any
trace of dust in Mkn 439 from a direct inspection of the $B$ band images on
a CCD frame. However, ellipse fitting analysis has been successfully
employed in the present study to infer the presence of dust in the inner regions of this galaxy
based on
multiband observations. 
The other peak occurs at 5\arcsec\ which corresponds to the brightest
region seen in
H$\alpha$. The depth of the dip between the two peak reduces at
longer wavelengths. The first peak and the dip is probably due
to dust while the second one corresponds to the blue region at the end of the bar.
Both these factors, namely dust and star forming regions contribute the maximum at shorter wavelengths.
At longer wavelengths,
the effects of both dust and the star forming regions are reduced hence
we see the underlying old stellar population. As a result the depth of the dip is reduced at
longer wavelengths.
Beyond 5\arcsec, the ellipticity starts dropping and reaches a value
($\sim$0.05) at 15\arcsec\ and remains
at a low value beyond that in all filters. Between 5\arcsec\ and 15\arcsec , the
isophotes at shorter wavelengths are rounder than the corresponding isophotes
at longer wavelengths indicating the presence of dust in this region of Mkn 439.
The position
angle is nearly constant in the inner 10\arcsec.
The luminosity profiles show an inner steeply rising part and an outer
exponential disk.
We derived the scale lengths of Mkn 439 in each of the filter bands.
This was done by marking the disk and fitting an exponential to the
surface brightness profile in this region. The range of fit was taken to be
from
18\arcsec\ to the region where the signal falls to 2$\sigma$ of the background.
The fit to the $H$ band is shown in Fig.\ref{fit}.  The scale lengths
derived were 0.97$\pm 0.14$ kpc in $B$, 0.84$\pm 0.02$ kpc in $R$
and 0.61$\pm 0.03$ kpc in $H$ band.

\section{Conclusions}
\begin{enumerate}
\item Mkn 439 is a peculiar galaxy made up of three distinct components: an
elliptical structure in the inner regions, a smooth outer envelope in
which this structure is embedded and a bar.
We detect massive star formation along the bar in Mkn 439.
This bar is misaligned with the main body of the galaxy.
\item The signature of the bar gets progressively fainter at longer wavelengths.
\item The stars in the bar are young and have not yet started influencing the
light in the near infrared region. This indicates that the galaxy has undergone
some perturbation which trigerred the bar formation and the starburst along the
bar in recent times.
\item There are indications for the presence of dust in the inner
15\arcsec\ of the galaxy.
\end{enumerate}
\begin{acknowledgements}
We are grateful to the anonymous referee for useful suggestions. One of the authors A. Chitre wishes to thank Tommy Wiklind for useful
discussions. The authors are thankful to Dr. K.S. Baliyan for helping with
observations. This work was supported by the Department of Space, Government
of India.
\end{acknowledgements}


\begin{thebibliography}{}
\bibitem[1983] {balz} Balzano V.A., 1983, ApJ 268, 602
\bibitem[1998] {bek} Bekki Kenji, 1998, ApJ 502, L133
\bibitem[1987]{bend} Bender R., M\"{o}llenhoff C., 1987, A\&A 177, 71
\bibitem[1999] {chitre} Chitre A., 1999, Ph.D thesis, Gujarat University
\bibitem[1999] {cucj} Chitre A., Joshi U.C., 1999, A\&AS {\it in press}
\bibitem[1994]{combes} Combes F., 1994, in: {\it The Formation and Evolution of Galaxies}, V Canary Islands Winter School of Astrophysics, eds. C. Mu\~{n}oz-Tu\~{n}on \& F. S\'{a}nchez, Cambridge Univ. Press, p.359
\bibitem[1987] {deut} Deutsch L.K., Willner S.P., 1987, ApJS, 63, 803
\bibitem[1989] {dev} Devereux N.A., 1989, ApJ 346, 126
\bibitem[1990] {ish} Ishizuki S., Kawabe R., Ishiguro M., et al., 1990, Nature 344, 224
\bibitem[1987]{jedr} Jedrejewski R.I., 1987, MNRAS 226, 747
\bibitem[1999]{jetal} Joshi U.C., et al., 1999 {\it in preparation}
\bibitem[1997]{jung} Jungweirt B., Combes F., Axon D.J., 1997, A\&AS 125, 497
\bibitem[1992]{land} Landolt A.U., 1992, AJ 104, 340
\bibitem[1984]{lo} Lo K.Y., Berge G.L., Claussen M.J., et al., 1984 ApJL 282, 59
\bibitem[1992a]{prietoa} Prieto M., Beckman J.E., Cepa J., et al., 1992a, A\&A 257, 85
\bibitem[1992b]{prietob} Prieto M., Longley D.P.T., Perez E., et al., 1992b, A\&AS 93, 557
\bibitem[1998]{rudrix} Rudnick G., Rix H., 1998, AJ, 116, 1163
\bibitem[1987]{soif} Soifer B.T., Sanders D.B., Madore B.F., et al., 1987, ApJ 320, 238
\bibitem[1998]{usui} Usui T., Saito M., Tomita A., 1998, AJ 116, 2166
\bibitem[1990]{van} van den Bergh S., Pierce M.J., 1990, ApJ 364, 444
\bibitem[1990] {wikthes} Wiklind T., 1990, Ph D thesis, Chalmers University of Technology, Sweden
\bibitem[1989] {wik} Wiklind T., Henkel C., 1989, A\&A 225, 1
\bibitem[1995]{woz} Wozniak H., Friedli D., Martinet L., et al., 1995, A\&AS 111, 115
\end{thebibliography}
\end{document}